\documentclass[aps,superscriptaddress,amsfonts,amssymb,amsmath,showpacs,nofootinbib, twocolumn]{revtex4}
\usepackage[dvips]{graphicx}

\begin{document}

\title{No asymptotically highly damped quasi-normal modes without horizons?}

\author{Cecilia Chirenti} 
\email{e-mail: cecilia.chirenti@ufabc.edu.br} 
\address{Centro de Matem\'atica, Computa\c c\~ao e Cogni\c c\~ao, UFABC, 09210-170 Santo Andr\'e, SP, Brazil}
\author{Alberto Saa}
\email{e-mail: asaa@ime.unicamp.br}
\address{
Departamento de Matem\'atica Aplicada,
Universidade Estadual de Campinas,
13083-859 Campinas,  SP, Brazil}
\author{Jozef Sk\'akala}
\email{e-mail: jozef.skakala@ufabc.edu.br}
\address{Centro de Matem\'atica, Computa\c c\~ao e Cogni\c c\~ao, UFABC, 09210-170 Santo Andr\'e, SP, Brazil}

\begin{abstract}
We explore the question of what happens with the asymptotically highly damped quasi-normal modes ($\ell$ fixed, $|\omega_{I}|\to\infty$) when the underlying  spacetime has {no} event horizons.  We consider the characteristic
oscillations of a scalar field in a large class of asymptotically flat spherically
symmetric static spacetimes without (absolute) horizons, such that the class accommodates the cases that are known to be of some sort of physical interest. The question of the asymptotic quasi-normal modes in such spacetimes 
is relevant to elucidate the connection between the behavior of the asymptotic 
quasi-normal modes and the quantum properties of event horizons, as put forward in some recent important conjectures.  We prove for a large class of asymptotically flat spacetimes without horizons that  the scalar field asymptotically highly damped modes do  {not} exist. This provides in our view additional evidence that there is indeed a close link between the asymptotically highly damped modes and the existence  of spacetime horizons (and their properties).
\end{abstract}


\maketitle

\section{Introduction}
It is a known result \cite{Wald1, Wald2} that in generic \emph{static} spacetimes (globally hyperbolic, or not) one can always define (for reasonable enough initial data) a sensible time evolution of a scalar field represented by a self-adjoint operator on a suitable Hilbert space. Many important features of a typical scattering can be described by a set of characteristic complex frequencies ($\omega=\omega_{R}+i \omega_{I}$), the quasi-normal modes (QNMs). In spherically symmetric spacetimes the quasi-normal modes
are labelled by the wave mode number $\ell$ and a discrete number $n$, such that $n$ grows with the decreasing damping time of the modes. More than a decade ago, it was conjectured by Hod \cite{Hod} that, due to the Bohr's correspondence principle, the asymptotically highly damped modes (the wave mode numbers fixed and $|\omega_{I}|\to\infty$) might carry important information about the quantum properties of the black hole horizon. The original conjecture of \cite{Hod} was modified by Maggiore \cite{Maggiore}, but the essence of the conjectures remains the same. The conjecture of \cite{Maggiore} was succesfully used in the case of many black hole spacetimes, (for the spherically symmetric black hole spacetimes see for example \cite{Skakala}). 

Furthermore, one can generically observe that in case of static, spherically symmetric \emph{black hole} spacetimes the asymptotically highly damped modes always exist and fulfil certain general patterns \cite{Das1, Das2, Kunstatter, Visser}. This fact can be seen as a strong support for the conjecture of Maggiore. Thus one might be interested in the question of what is going to happen with the asymptotic highly damped modes in static spherically symmetric spacetimes in case there are \emph{no} (absolute) horizons. It is very intuitive to expect that from the point of view of the conjectures in question \cite{Hod, Maggiore}, such asymptotically highly modes might \emph{not} exist. The non-existence of such modes was shown in one particular case of a spacetime without horizon, in the case of a Reissner-Nordstr\"om naked singularity \cite{Chirenti}. Thus the result of \cite{Chirenti} can be seen as supporting the conjectures that the QNMs might generally relate to the spacetime horizons in an essential way\footnote{Another example of a spacetime without horizon is the pure anti-de-Sitter spacetime, where no quasi-normal modes exist, hence no asymptotic quasi-normal modes exist.}. The aim of this short paper is to generalize this result to large classes of asymptotically flat spherically symmetric static spacetimes \emph{without} horizons, accommodating all the known interesting cases. As a result of this fact, this paper should add much more evidence to support the intuition based on the mentioned conjectures. 

\section{Static spherically symmetric spacetimes without (absolute) horizons}

The metric of a general spherically symmetric static spacetime can be cast in the form
\begin{equation}
\label{metric}
ds^2 = -f(r)~dt^{2}+f^{-1}(r)~dr^{2}+h^2(r)~d\Omega^{2}.
\end{equation}
We assume $f(r), ~h'(r)$ being continuous and a.e 1-differentiable with bounded first derivatives everywhere outside infinitesimal neighbourhood of zero. Furthermore, since the spacetime has no event horizon, $f(r)$ is positive 
on $(0,\infty)$. In order to guarantee an  asymptotically flat spacetime, one
 needs the further restrictions $f(r) \to 1$ and $h(r) \sim r $ for $r\to \infty$. Under the geometry
 (\ref{metric}), the area of the spherical surface with radius $r$ is given by $4\pi h^2(r)$ and, hence, is natural also to assume $h(0)=0$ and $h'(r)>0$. 
 
The massless Klein-Gordon time-independent equation for the metric (\ref{metric}) can be cast, by introducing the usual tortoise-coordinate $x(r)$
\begin{equation}
\label{tort}
\frac{dr}{dx}= f(r),
\end{equation}
as
\begin{equation}
\label{eq1}
\left(\partial^{2}_{x} +\omega^2 -V(r)\right)\phi_{\ell}=0,  
\end{equation}
where $\psi_\ell(t,r)= \exp(-i\omega t) \phi_{\ell}(r)/h(r)$, 
 with $\psi_{\ell}$ standing for the spherical harmonic component of the scalar field, and
\begin{equation}
\label{pot}
V(r) = f(r) \left[\frac{\ell(\ell+1)}{h^2(r)} + \frac{1}{h(r)} \frac{d}{dr}\left(
f(r)\frac{d}{dr}h(r)
\right) \right].
\end{equation}
On general grounds, one can expect $V(r)\to 0$ for $r\to\infty$ due to the asymptotic 
flatness, and a diverging $V(r)$ for $r\to 0$ due to the centrifugal barrier (at least). 

Let us explore the poles of the scattering amplitude, such that do not belong to the bound states, the quasi-normal modes (QNMs). Some superposition of QNMs dominates the time evolution of an arbitrary perturbation at a given point within some specific time scale. In the scattering problem in a spacetime without horizons the QNMs are determined by the normal mode boundary condition at 0 and at the same time by purely outgoing radiation condition at infinity $\sim e^{i\omega x}$. The reason for this is the following: the Green function is composed from two solutions, such that each of them fulfils the given boundary condition at one of the ``ends''. (This is because one requires boundedness of the Green function in the spatial variables.) Then the poles of the Green function (QNMs) occur where the two solutions coincide, (for a more detailed argumentation see \cite{Chirenti}). (Note also that considering only scattering that is reflective one can immediately see that such two boundary conditions \emph{cannot} be fulfilled at the same time for \emph{real} frequencies $\omega$.) 

Unfortunately since the potential is typically non-compact and the frequencies $\omega$ have negative imaginary parts, (unless instabilities occur), one of the solutions is exponentially suppressed, whereas the other one corresponding to the ``outgoing'' wave is exponentially growing, and one has to give the ``outgoing radiation condition'' a clearer meaning. (If the potential had a compact support, one could claim the solutions to be directly proportional to the outgoing waves, but since the plane waves are only approximations to the solutions as one approaches infinity, one needs exponential precision in the error to rule out the ``incoming'' wave solution.) On the other hand the outgoing/incoming wave solutions behave as $A_{\pm}e^{\pm i\omega x}\left(1+O(1/x)\right)$, see \cite{Nollert}, so let us follow the suggestion of \cite{Motl}, analytically continue the solutions in the complex plane in $x$ and pick the purely outgoing radiation condition on the line $Im(\omega x)=0$ as $\omega x \to\infty$. (This means we pick the solution in the region where the two solutions are purely oscillatory  and of comparable sizes).

The asymptotically highly damped quasi-normal frequencies are with our normal mode convention characterized by  $-\omega_{I}\to \infty$ and $\ell$ fixed, implying that the highly damped quasi-normal modes should obey the approximate equation
\begin{equation}
\label{eq2}
\left(\partial^{2}_{x} +\omega^2  \right)\phi_{\ell}=0
\end{equation}
everywhere apart from a small neighbourhood of $r=0$. Of course, the two linearly independent solutions of (\ref{eq2}) are plane waves:
\begin{equation}
\label{sol1}
\phi_{\ell} = C_1 e^{i\omega x} + C_2 e^{-i\omega x}.
\end{equation}
The pure outgoing mode in the infinity then corresponds to 
$\psi_\ell(t,r) \sim   e^{-i\omega (t-x(r))}/r $, hence it is picked by $C_2=0$. For $|\omega| \gg 1$, the potential (\ref{pot}) will effectively affect the dynamics of the scalar field only in a region near $r=0$. The crucial point in our analysis is that the dynamics near $r=0$ can be solved and will determine the constants $C_1$ and $C_2$ for $|\omega| \gg 1$ in a way that will prevent the appearance of the quasinormal oscillations. (This is because one can approximate the scalar field equation everywhere by the form of the equation near zero, since in the region in which the approximation of the potential near zero ceases to hold, one can neglect the potential as a whole with respect to the $\omega^{2}$ term.) Relaxing the condition $|\omega| \gg 1$, the solution (\ref{sol1}) will be a good approximation for (\ref{eq1}) only for large $r$, and the natural boundary condition selected by the dynamics near $r=0$ can be translated in a $\omega$-dependent relation between the coefficients $C_1$ and $C_2$, giving rise to a characteristic equation $C_2(\omega)=0$ for the quasinormal modes.

We have basically two different kinds of interesting spacetimes in our analysis: spherically symmetric stellar spacetimes ($f(0)$ finite and positive) and naked singularities 
($f(r)\sim r^{-\epsilon}$, $\epsilon\neq 0 ~\wedge~ \epsilon >-1$, for $r\to 0$). We will consider them separately, although the analyses are similar. Some of the cases corresponding to $f(0)=0$ will be also left for the concluding section. 

\subsection{Stellar spacetimes}
 
The choices $f(r) \sim A + Br^\epsilon$ and $h(r) \sim Dr^\delta$ for $r\to 0$, with positive $A,D,\epsilon$, and
$\delta$, seem generic enough to accommodate all known stellar spacetimes. In fact, there are further restrictions among such constants. The divergent terms of the Ricci scalar for such choices of 
$f(r)$ and $h(r)$ read 
\begin{equation}
\label{scurv}
R \sim  \frac{2}{(D r^{ \delta})^2} - 2\delta(3\delta-2)\frac{A}{r^2}  ,
\end{equation}
for $r\to 0$. The only way to assure $R$ finite for $r\to 0$ is demanding 
$\delta =1$ and $A=D^{-2}$. (The stellar spacetimes must ensure that also the other curvature invariants will be finite, but we will perform more general analysis applying only the given Ricci curvature finiteness condition, which automatically contains regular stellar spacetimes as its subcase.) With this values, the potential (\ref{pot})
is given by
\begin{equation}
V(r) \sim \frac{\ell(\ell+1)}{D^4r^2} + \frac{\epsilon B}{D^2r^{2-\epsilon}} 
\end{equation}
for $r\to 0$. Notice that the condition of $f(0) = D^{-2}$ finite assures from (\ref{tort}) that ~$x \sim D^2 r+C$~ for $r\to 0$. Let us further choose $C=0$, hence $x(0)=0$. Since $\epsilon >0$, the dominant term
of the potential near the origin is the centrifugal barrier
\begin{equation}
\label{pot2}
V(x) = \frac{\ell(\ell+1)}{x^2}.  
\end{equation}
The general solution of (\ref{eq1}) with the potential (\ref{pot2}) is given in terms
of Bessel functions
\begin{equation}\label{solutions}
 \phi(x) = \sqrt{\omega x} \left(C_3  ~J_{\beta}(\omega  x)+C_4  ~J_{-\beta}(\omega  x)\right)  ,
\end{equation}
where in this case $\beta = \ell+1/2$~ and is a non-integer.  
To impose the boundary condition at $Im(\omega x)=0$, we understand the solution \eqref{solutions} to be analytically continued into the complex plane in $\omega\cdot x$. Since the solutions are multivalued functions of the complex variable $\omega\cdot x$ around zero, their analytical extensions require taking convenient branch cut. The branch cut can and will be taken along the half-line $\omega\cdot x$ real and negative.

Notice that the regularity of the scalar field
 $\psi_\ell$ at the origin and the properties of the Bessel functions close to zero automatically select the reflection condition $\phi_{\ell}(0)=0$,  which corresponds to $C_4=0$. The solution picked at the origin then behaves for $\omega x \to \infty$ as 
\begin{eqnarray}\label{asymptotic}
\sqrt{\omega x}\cdot J_{\ell+1/2}(\omega x)\approx ~~~~~~~~~~~~~~~~~~~~~~~~~~~~~~~~~~~~ 
\\ \sqrt{\left(\frac{2}{\pi}\right)}~ \cos\left(\omega x-\frac{(\ell+1)\cdot\pi}{2}\right)+O\left(\frac{1}{x}\right).\nonumber
\end{eqnarray}
 This means the reflection condition imposed on
(\ref{sol1}) implies $C_1=\pm C_2$, leaving no room for the fulfilment of the highly damped quasinormal mode condition $C_2=0$.

\subsection{Naked singularities}

In this case one has $f(r) = B r^{-\epsilon}$ ~and $h(r) = Dr^\delta$~ for $r\to 0$ with positive $B, D, \delta$,~ and $\epsilon>-1, ~\epsilon\neq 0$. (Again, such conditions should be fulfilled for a fairly generic naked singularity and such a metric is of a Szekeres-Iyer form \cite{Szekeres}.) The tortoise coordinate in this case is given near the radial center as ~$x=[B(\epsilon+1)]^{-1}r^{\epsilon+1}$~(again by setting $x(0)=0$) and the potential $V(x)$ has near zero two terms:
\begin{equation}
V(x) = \frac{P_1}{ x^{\mu}}
+ \frac{P_2}{x^{2}},
\end{equation}
where $\mu =\frac{2\delta+\epsilon }{\epsilon+1}$, $P_1=\frac{\ell(\ell-1)}{D^{2}[B(\epsilon+1)]^\mu }$, and $P_2=-\frac{\delta}{ \epsilon+1 }\left(- \frac{\delta}{  \epsilon+1   }+1\right)$.
In case $\mu\leq 2$ the potential again behaves close to zero as $\sim x^{-2}$. This case corresponds to $\epsilon/2+1\geq \delta$. In the case $\mu>2$ the scalar field equation has an irregular singular point at the radial center and much less can be said about the solutions. We will consider only the case $\mu\leq 2$, but this case already includes all the well known naked singularity solutions. For $\mu\leq 2$, unless $\beta$ is integer, the solutions are again given by the equation \eqref{solutions}. In case $\beta$ is an integer, the solutions are given as 
\[\sqrt{\omega x}~[C_{3}~J_{\beta}(\omega x)+C_{4}~Y_{\beta}(\omega x)],\]
where $Y_{\beta}$ is a Bessel function of the second kind.
The parameter $\beta$ is given in the naked singularity case for $\mu<2$ as: 
\[\beta=\left|\frac{1}{2}-\frac{\delta}{\epsilon+1}\right|,\]
and for $\mu=2$ and $P_1+P_2 >-1/4$ as: 
\[\beta=\sqrt{1/4+P_{1}+P_2}.\]

To determine the quasinormal modes one has to first select the normal modes at the origin, such that $\psi_{\ell}(0)$ is finite. This can be uniquely done for $\delta\geq (\epsilon+1)/2$, which means in such case the time evolution of the scalar field is \emph{unique}. In case the singularity is too strong and $\delta < (\epsilon+1)/2$, both of the solutions $\psi_{\ell}$ are convergent at zero and the time evolution is non-unique. (This means the singularity has ``hair'' and the quasinormal modes depend on the ``hair''.) In such case one can still uniquely choose normal modes such that fulfil $\psi_{\ell}(0)=0$, corresponding to a reflective boundary condition at the singularity. This represents a preferred choice for the time evolution, in the language of operators corresponding to what is called a Friedrich's self-adjoint extension of a symmetric operator. In case the time evolution is non-unique we will automatically impose such a most natural choice of time evolution.
So in each of the cases let us proceed exactly as before and impose the $\psi_{\ell}(0)=0$ condition at the radial center which leads (for both $\beta$ integer, or non-integer) to the condition $C_4=0$. But by using the asymptotic behavior at $\omega x\to\infty$ given by the equation \eqref{asymptotic}, the choice of normal modes leads, exactly as before, to a non-trivial linear combination of outgoing and incoming waves, such that it is independent of $\omega$. This means one cannot fulfil at the same time the boundary conditions at both ``ends''.
(For the sake of completeness, let us say that one can repeat the same analysis with the same results also for subcases of the case $\epsilon=0$ and $\delta \neq 1$.)   

 Let us mention three types of naked singularities of some interest, such that fall under our analysis: 
 The negative mass Schwarzschild singularity, the vacuum solution of the Einstein equations, has the parameters given as $\epsilon=\delta=1$. This means $\mu=3/2<2$ and $\delta/(\epsilon+1)=1/2$, which means the time evolution is in this case unique. The Reissner-Nordstr\"om naked singularity arising from a coupling between gravity and electromagnetic field has parameters given as $\epsilon=2$ and $\delta=1$, which means $\mu=4/3<2$ and $\delta/(\epsilon+1)=1/3<1/2$. This means in the case of Reissner-Nordstr\"om naked singularity the time evolution is non-unique. The Wyman's \cite{Wyman} solution that provides a naked singularity arising from a minimal coupling of gravity to a charged massive scalar field has,  ($0<\alpha<1$), $\delta=(1-\alpha)/2$ and $\epsilon=-\alpha$. Here $\alpha$ is simply related to the scalar field mass and the scalar field charge. This means $\mu=1-\alpha/(1-\alpha)<2$ and $\delta/(\epsilon+1)=1/2$. Thus in case of Wyman solution the time evolution is unique. Let us mention here that the same results concerning the uniqueness of the scalar field time evolution were obtained for these three naked singularities via the language and methods of functional analysis in \cite{Ishibashi}. (The paper \cite{Ishibashi} provides also more details about the naked singularities in question.)

\section{Conclusions}

Note that the case $\epsilon\le -1$, for $f(r)\sim r^{-\epsilon}$ as $r\to 0$, is qualitatively different from  the cases analysed before, since one has $x(r)\to -\infty$
 for $r\to 0$. In this case the light rays reach the radial center in the infinite coordinate time. One cannot explore the asymptotic properties of the solutions of (\ref{eq1}) to select the natural boundary conditions at $r=0$ in the same way and we indeed may have room for the asymptotically highly damped quasinormal modes. An infinitesimal mass Schwarzschild black hole belongs to this class, for instance.

In this paper we generalized our previous result from \cite{Chirenti} to,  (at least), large classes of spherically symmetric asymptotically flat static spacetimes \emph{without} (absolute) horizons. Our results provide evidence that for the class of spacetimes in question the asymptotically highly damped modes ($\ell$ fixed and $|\omega_{I}|\to\infty$) do \emph{not} exist. This means the non-existence of (absolute) spacetime horizons could be conjectured to mean the non-existence of the asymptotic modes. As mentioned in the introduction, it is widely observed \cite{Das1, Das2, Kunstatter, Visser}, that the existence of (absolute) spacetime horizons leads to the existence of the asymptotic quasi-normal modes. One can then \emph{conjecture} that the equivalence: \emph{existence of (absolute) spacetime horizons} $\leftrightarrow$ \emph{existence of asymptotically highly damped modes} might be completely general, confirming the intuition one might have from the currently popular conjectures linking the modes to the horizon properties. To our knowledge, our result covers already all the static spherically symmetric (asymptotically flat) spacetimes \emph{without} horizons, that are of some sort of interest: the (regular) stellar spacetimes of the given type and the most ``famous'' static naked singularity solutions. 
Note that in case of spacetimes of a static neutron star the computed behaviour of the imaginary parts of the (axial) w-modes (for the fixed $\ell$) already suggests  (see some of the plots of \cite{Kokkotas}) that the absolute values of the imaginary parts might be upper bounded. 

Let us also add that models of a theoretical interest from the point of view of the topic analysed in this paper are gravastars. They behave close to the radial center as the de-Sitter spacetime \cite{Rezzolla}, hence fall in the $F=G=0$ case of this paper, and the quasi-normal mode spectrum for fixed $\ell$ has to be bounded in the imaginary part. One might be interested in what happens if a sequence of static gravastar solutions approaches the black hole horizon, (that is, a sequence of mass $M$ gravastars with surface radius
$R \to 2M^{+}$). One of the possibilities is that the upper bound of the ($\pm$) quasi-normal frequency imaginary part grows to infinity as the horizon is approached, the other possibility is that there will be a sudden discontinuous qualitative change once the black hole horizon is reached. (Such a qualitative change might be seen as a form of ``phase transition''.)

\medskip

{\bf Acknowledgments:}  This research was supported by FAPESP, CNPq and the Max Planck Society.


\begin{thebibliography} {99}

\bibitem{Wald1}
R.M. Wald, \emph{Dynamics in nonglobally hyperbolic, static spacetimes}, J.Math.Phys {\bf 21}, 2802, 1980

\bibitem{Wald2}
A. Ishibashi and R.M. Wald, \emph{Dynamics in Non-Globally-Hyperbolic Static Spacetimes II: General Analysis of Prescriptions for Dynamics}, Class.Quant.Grav. {\bf 20} (2003) 3815-3826, arXiv:gr-qc/0305012 

\bibitem{Hod}
 S. Hod, \emph{Bohr's correspondence principle and the area spectrum of
                        quantum black holes}, Phys.Rev.Lett. {\bf 81},  4293, 1998, arXiv:gr-qc/9812002,

\bibitem{Maggiore}
M. Maggiore, \emph{The Physical interpretation of the spectrum of black
                        hole quasinormal modes}, Phys.Rev.Lett. {\bf 100}, 141301, 2008, arXiv:gr-qc/0711.3145

\bibitem{Skakala}
J. Skakala, \emph{Quasinormal modes, area spectra and multi-horizon
                        spacetimes}, JHEP {\bf 1206}, 094, 2012, arXiv:gr-qc/1204.3566
 
\bibitem{Das1}
S. Das and S. Shankaranarayanan, \emph{High frequency quasi-normal modes for black-holes with generic singularities}, Class.Quant.Grav.{\bf 22}:L7, 2005, arXiv:hep-th/0410209 

\bibitem{Das2}
A. Ghosh, S. Shankaranarayanan and  S. Das, \emph{High frequency quasi-normal modes for black holes with generic singularities II: Asymptotically non-flat spacetimes}, Class.Quant.Grav. {\bf 23} (2006) 1851-1874, arXiv:hep-th/0510186 

\bibitem{Kunstatter}
R.G. Daghigh and G. Kunstatter, ~\emph{Highly Damped Quasinormal Modes of Generic Single Horizon Black Holes}, Class.Quant.Grav. {\bf 22} (2005) 4113-4128, arXiv:gr-qc/0505044 

\bibitem{Visser} 
A.J.M. Medved, D. Martin and M. Visser, ~\emph{Dirty black holes: Quasinormal modes},
Class.Quant.Grav. {\bf 21}, (2004) 1393-1406, arXiv:gr-qc/0310009


\bibitem{Chirenti}
C. Chirenti, A. Saa and J. Skakala, \emph{Quasinormal modes for the scattering on a naked Reissner-Nordstrom singularity}, arXiv:gr-qc/1206.0037

\bibitem{Szekeres}
P. Szekeres and V. Iyer, \emph{Spherically symmetric singularities and strong cosmic censorship}, Phys. Rev. D {\bf 47}, 4362 ~1993,


\bibitem{Wyman}
M. Wyman, \emph{Static spherically symmetric scalar fields in general relativity}, Phys. Rev. D{\bf 24}, 839, 1981

\bibitem{Ishibashi}
A. Ishibashi and A. Hosoya, \emph{Who's afraid of naked singularities?},  Phys.Rev. D{\bf 60} (1999) 104028, arXiv:gr-qc/9907009

\bibitem{Nollert}
H.P. Nollert, \emph{Quasinormal modes: the characteristic `sound' of black holes and neutron stars}, Class. Quantum Grav. {\bf 16} R159, 1999,

\bibitem{Motl}
L. Motl and A. Neitzke, \emph{Asymptotic black hole quasinormal frequencies},  Adv.Theor.Math.Phys. {\bf 7} (2003) 307-330, arXiv:hep-th/0301173

\bibitem{Kokkotas}
K. Kokkotas and B.D.Schmidt, \emph{Quasi-Normal Modes of Stars and Black Holes}, LivingRev.Rel.2:2,1999, arXiv:gr-qc/9909058

\bibitem{Rezzolla}
C. Chirenti and L. Rezzolla, \emph{How to tell a gravastar from a black hole}, Class.Quant.Grav. {\bf 24}:4191-4206,2007, arXiv:0706.1513

\end{thebibliography}
\end{document}